\documentclass[3p, twocolumn]{elsarticle}

\usepackage{lineno,hyperref}
\modulolinenumbers[5]

\journal{Journal of Molecular Spectroscopy}

\def\alhp{AlH$^+$}
\def\XdS{X$^{2}\Sigma^{+}$}
\def\AdP{A$^{2}\Pi$}
\def\BdS{B$^{2}\Sigma^{+}$}
\def\CdS{C$^{2}\Sigma^{+}$}
\def\DdP{D$^{2}\Pi$}
\newcommand{\fig}[1]{Fig.\,\ref{#1}}

\newcommand{\Ba}{Ba$^+$}

\newcommand{\AlH}{AlH$^+$}

\newcommand{\Bae}{$^{138}$Ba$^+$}

\newcommand{\+}{$^+$}

\begin{document}

\begin{frontmatter}

\title{Rotational State Analysis of AlH$^+$ by Two-Photon Dissociation }

\author[NU]{Christopher~M~Seck}
\author[S1,S2]{Edward~G~Hohenstein}
\author[NU]{Chien-Yu~Lien}
\author[NU]{Patrick~R~Stollenwerk}
\author[NU]{Brian~C~Odom\corref{mycorrespondingauthor}}
\ead{b-odom@northwestern.edu}

\address[NU]{Department of Physics and Astronomy, Northwestern University\\ 2145 Sheridan Rd., Evanston, IL 60208}
\address[S1]{Department of Chemistry and the PULSE Institute,
Stanford University, \\ Stanford, CA 94305}
\address[S2]{SLAC National Accelerator Laboratory,\\ Menlo Park, CA 94025}

\begin{abstract}
We perform \textit{ab initio} calculations needed to predict the cross-section of an experimentally accessible ($1+1'$) resonance-enhanced multiphoton dissociation (REMPD) pathway in \AlH.  Experimenting on \AlH\ ions held in a radiofrequency Paul trap, we confirm dissociation via this channel with analysis performed using time-of-flight mass spectrometry.  We demonstrate the use of REMPD for rotational state analysis, and we measure the rotational distribution of trapped \AlH\ to be consistent with the expected thermal distribution.  \AlH\ is a particularly interesting species for ion trap work because of its electronic level structure, which makes it amenable to proposals for rotational optical pumping, direct Doppler cooling, and single-molecule fluorescence detection.  Potential applications of trapped \AlH\ include searches for time-varying constants, quantum information processing, and ultracold chemistry studies.
\end{abstract}

\begin{keyword}
multiphoton dissociation \sep molecular ion \sep ion trap \sep aluminum monohydride \sep AlH$^+$ \sep REMPD
\end{keyword}

\end{frontmatter}


\section{Introduction}
Because of their rich internal structure, trapped molecules offer great potential for precision spectroscopy experiments, including searches for time-variation of fundamental constants~\cite{shelkovnikov_stability_2008}, parity violation studies~\cite{demille_using_2008}, and searches for fundamental electric dipole moments~\cite{hudson_improved_2011, leanhardt_high-resolution_2011, baron_order_2014}.  Trapped polar molecules have also been proposed for use in quantum information processing~\cite{demille_quantum_2002, schuster_cavity_2011}.  Trapped neutral and charged molecules are also excellent candidates for ultracold chemistry studies and coherent control applications.

Generally, rovibrational state control and readout present a challenge for all these applications.  In light of these challenges, it is interesting to consider a special class of ``alkali-like" molecular ions. We have proposed that these species can quickly be cooled to their rovibrational ground state with a single spectrally filtered pulsed laser~\cite{lien_optical_2011}, detected by fluorescence at the single-molecule level, and in some cases directly Doppler cooled~\cite{nguyen_challenges_2011,nguyen_prospects_2011}.

Here we study one such molecular ion, \AlH.  Owing largely to astrophysical interest, some spectroscopic properties of the \AlH\ molecular ion are already well understood with good agreement between experiment~\cite{almy_band_1934, muller_chemiluminescent_1986, muller_spectroscopic_1988, szajna_high-resolution_2011} and theory~\cite{klein_ab_1982, nguyen_challenges_2011}.  Here, we extend our previous calculations~\cite{nguyen_challenges_2011} to include states relevant for an experimentally convenient pathway for REMPD rotational state readout, and we predict an experimentally accessible dissociation cross-section.

We also perform experiments on Coulomb crystals of \AlH\ held in a radiofrequency (RF) Paul trap and sympathetically cooled via laser cooling of co-trapped \Ba.  Using time-of-flight mass spectrometry (TOFMS), we make the first demonstration of REMPD of \AlH, and we use REMPD/TOFMS for rotational state analysis of our trapped sample. Rotational state analysis by REMPD has previously been used for HD\+~\cite{roth_rovibrational_2006} and MgH\+~\cite{roth_rovibrational_2006}, where the molecular fraction dissociated was analyzed by observing deformation of a fluorescing Coulomb crystal of co-trapped atomic ions.  Also in the current issue of this journal~\cite{ni_state-specific_2014}, Ni \textit{et al.} analyze the rotational distribution of non-crystallized trapped HfF\+\ using REMPD/TOFMS.

\section{Computational Details}

Potential energy curves and transition dipole moments for \alhp\ were
computed with the multireference singles and doubles configuration
interaction (MRSDCI) method. All computations were performed with the {\tt
MOLPRO} quantum chemistry package~\cite{MOLPRO}. The molecular orbitals used
in the MRSDCI computations were obtained from the state-averaged complete
active space self-consistent field (SA-CASSCF)
method.\cite{RuedenbergIJQC1979,RoosCP1980} The SA-CASSCF orbitals
were optimized with an active space consisting of three electrons in the
$4\sigma2\pi_x2\pi_y5\sigma6\sigma$ orbitals while averaging over the lowest
four electronic states. In the MRSDCI wavefunction, the 1s Al orbital was
constrained to always be doubly occupied. The aug-cc-pCVQZ basis set
was used for all computations.\cite{Dunning:1989:1007, Woon:1993:1358}

\begin{figure}[htbp!]
  \centering
  \includegraphics*[width=3in]{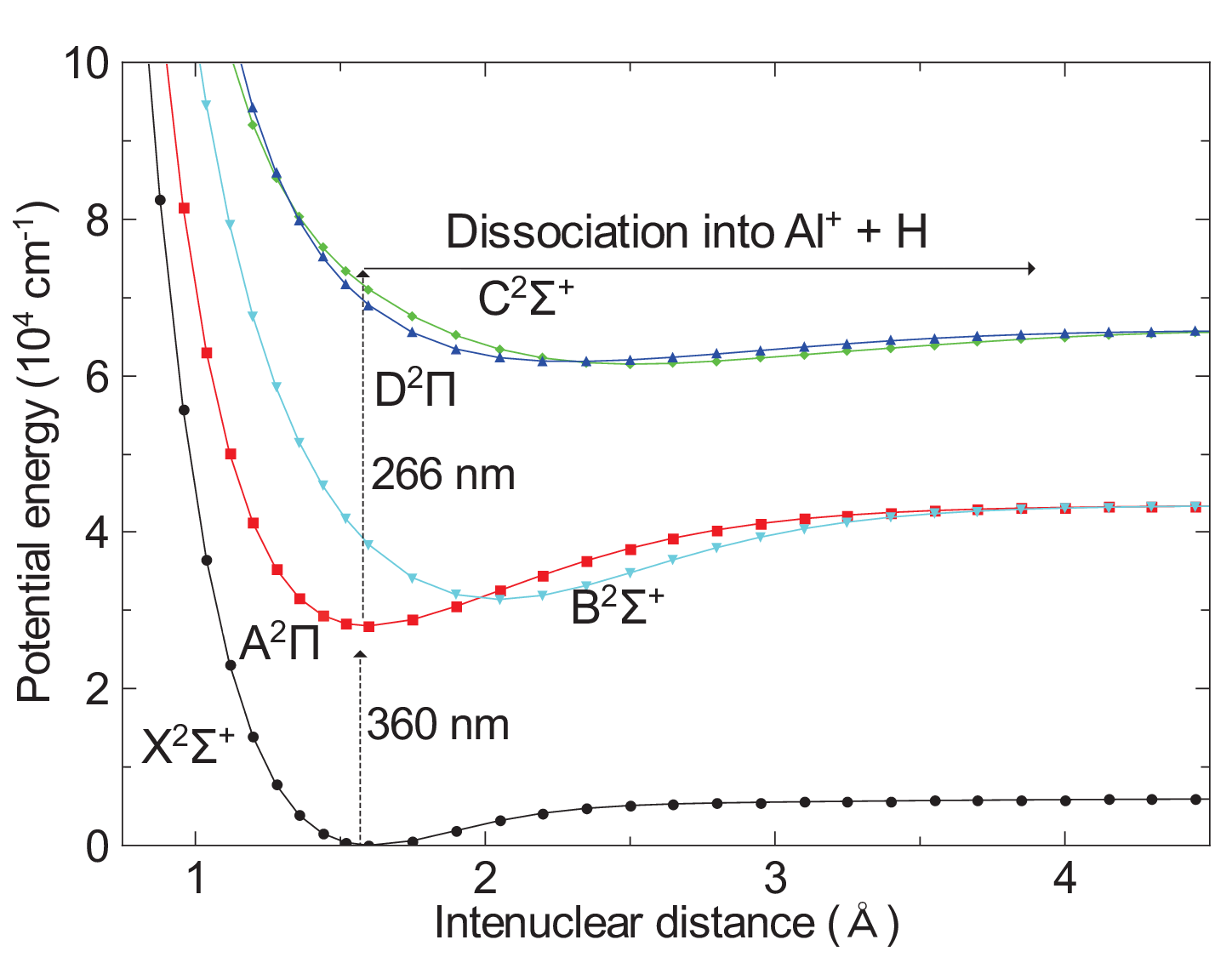}
  \caption{Calculated potential energy surfaces of the lowest several electronic states of \AlH. The resonant-enhanced two-photon dissociation pathway used in the experiment is also shown.}
  \label{PEC}
\end{figure}

\begin{figure}[htbp!]
  \centering
  \includegraphics*[width=3in]{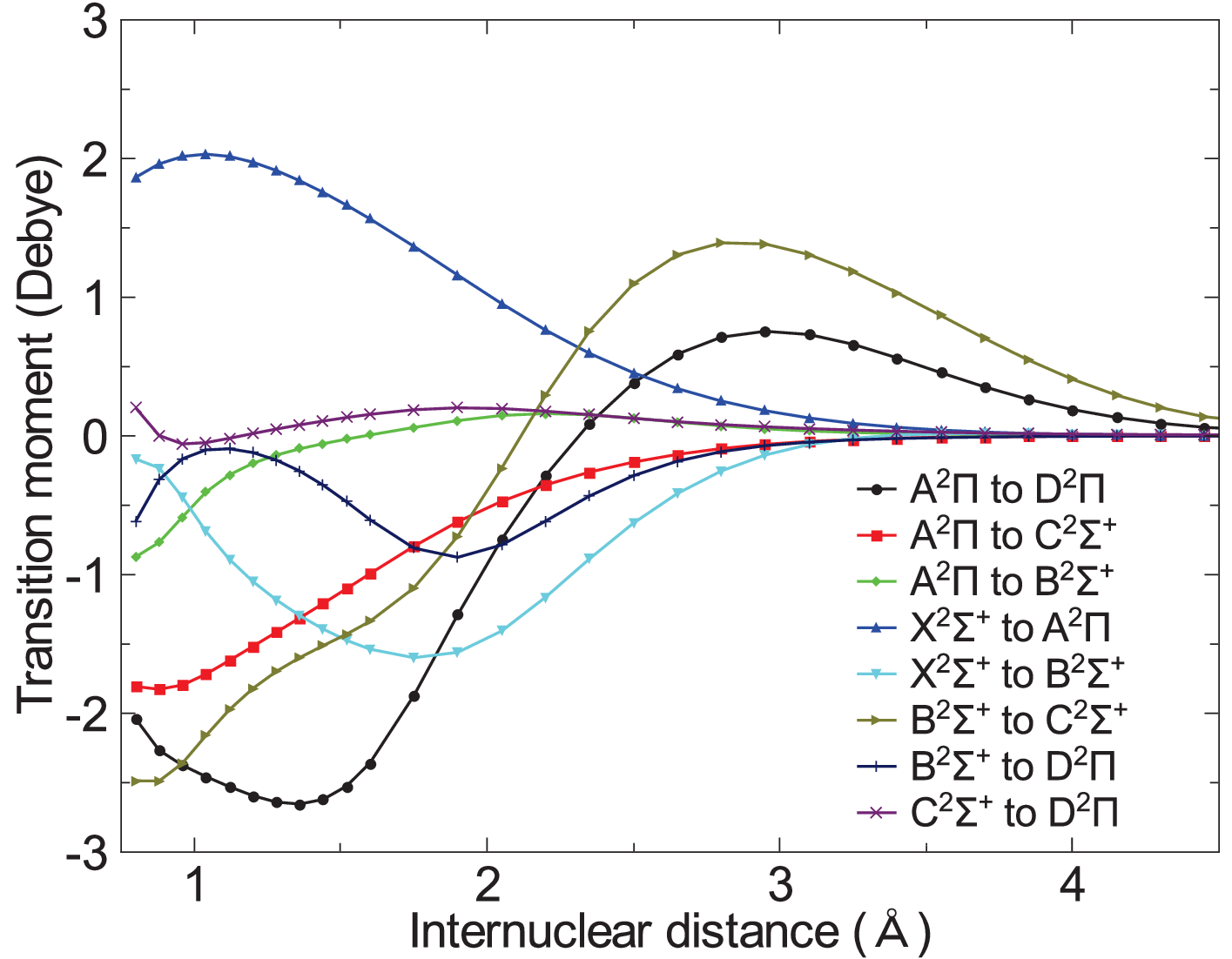}
  \caption{Calculated transition dipole moment functions of selected electronic transitions in \AlH.}
  \label{TDM}
\end{figure}

In a previous study~\cite{nguyen_challenges_2011}, we found the results obtained from MRSDCI to be in excellent agreement with approximate coupled-cluster singles, doubles and
triples (CC3)~\cite{Christiansen:1995:7429, Christiansen:1995:75} as well
as the experimentally determined bond length and vibrational frequency of \alhp~\cite{muller_spectroscopic_1988}. In the present work, the \CdS\
and \DdP\ states could not be studied with CC3, since current
implementations are not fully spin-adapted ({\em i.e.} CC3 wavefunctions
are not proper spin eigenfunctions). This led to spurious mixing of
the two higher-lying doublet states with quartet states. For this reason, the
present work relies on MRSDCI computations exclusively. A drawback of this approach
is that the energy spacings at dissociation do not match experimental
results. In order to fix the energy spacings, constant shifts were
applied so that, at dissociation, the gap between the \XdS and \AdP\//\BdS\
is 266.9157 nm and \XdS and \CdS\//\DdP\ states is 167.0787
nm~\cite{Martin:1979:817}. We expect that the nonparallelity errors of the MRSDCI method are small, and the application of a small constant shift is reasonable.  Calculated potential energy surfaces and transition dipole moment functions are shown in Figs.~\ref{PEC} and ~\ref{TDM}.

\begin{figure}[htbp!]
  \centering
  \includegraphics*[width=3in]{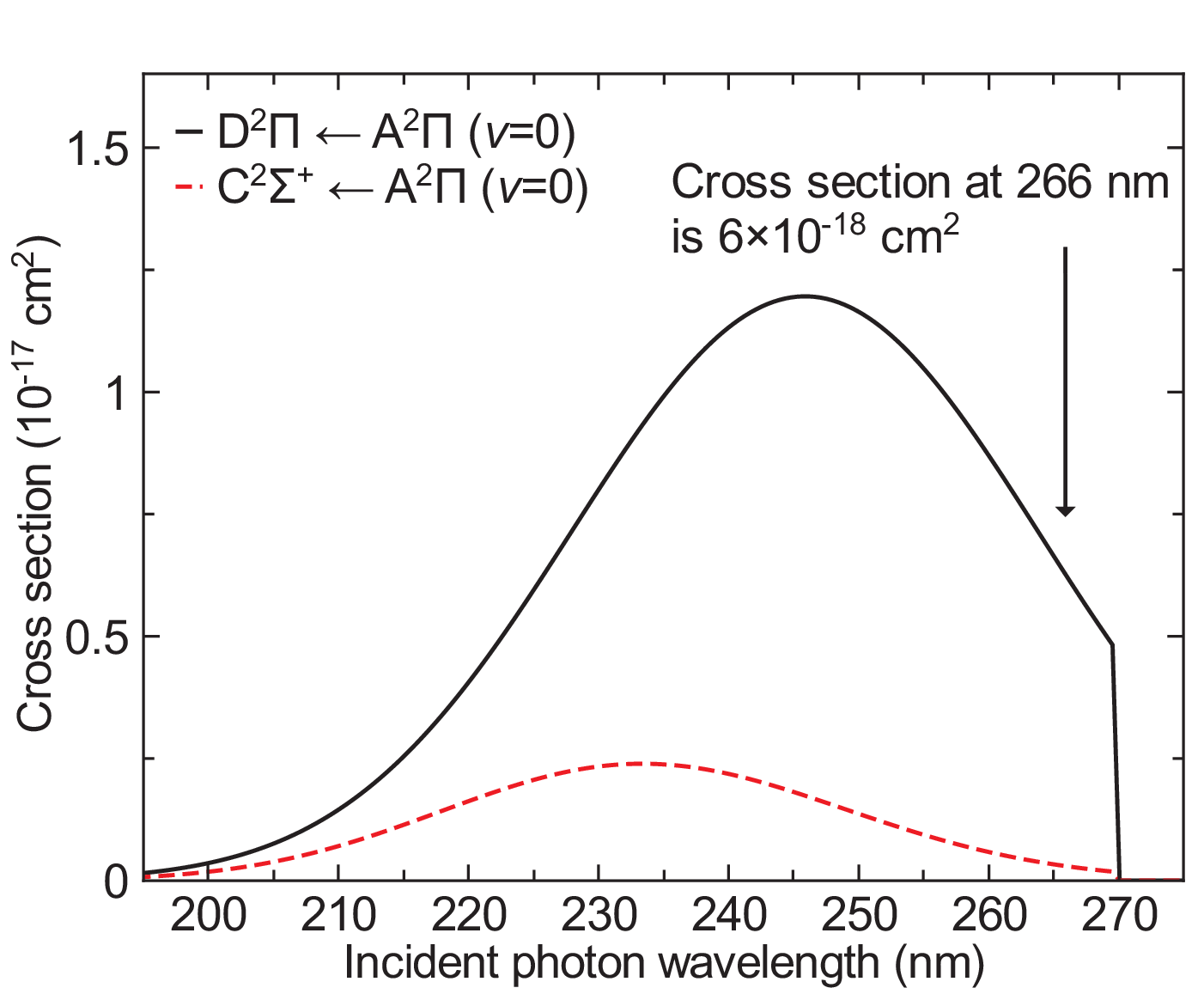}
  \caption{Calculated photodissociation cross-sections from the \AlH\ A$^{2}\Pi$, $v=0$ state. The sharp cutoff near 270 nm corresponds to a photon energy insufficient to excite to the continuum.}
  \label{cross_section}
\end{figure}

Photodissociation cross-sections were calculated using the program BCONT 2.2~\cite{BCONT}, using as input our calculated surfaces and transition moments. In Fig. \ref{cross_section}, the experimentally relevant photodissociation cross-sections from the A$^{2}\Pi$, $v=0$ state are plotted as a function of photon wavelength. Experimentally, as shown in Fig.~\ref{PEC}, 266 nm light is used for the dissociating photon. At this wavelength the cross-section is dominated by the D$^{2}\Pi \leftarrow A^{2}\Pi$, $v=0$ transition and is calculated to be $\sigma_0 = 6\times10^{-18}$ cm$^{2}$.

Experimentally, it could also be convenient to use 532 nm light to dissociate from A$^{2}\Pi$ to the B$^{2}\Sigma^+$ continuum.  However, even though this cross-section is near its maximum at this wavelength, the calculated cross-section of $3\times10^{-22}$ cm$^{2}$ is inconveniently small. Another experimental consideration is whether the 266 nm light might drive bound-bound transitions from X$^{2}\Sigma^+$ to A$^{2}\Pi$ or B$^{2}\Sigma^+$.  Although such processes are energetically allowed, the vibrational overlap is expected to be small, and we see no experimental evidence for such transitions.

\section{Experimental Details}

\begin{figure}[htbp!]
  \centering
  \includegraphics*[width=3in]{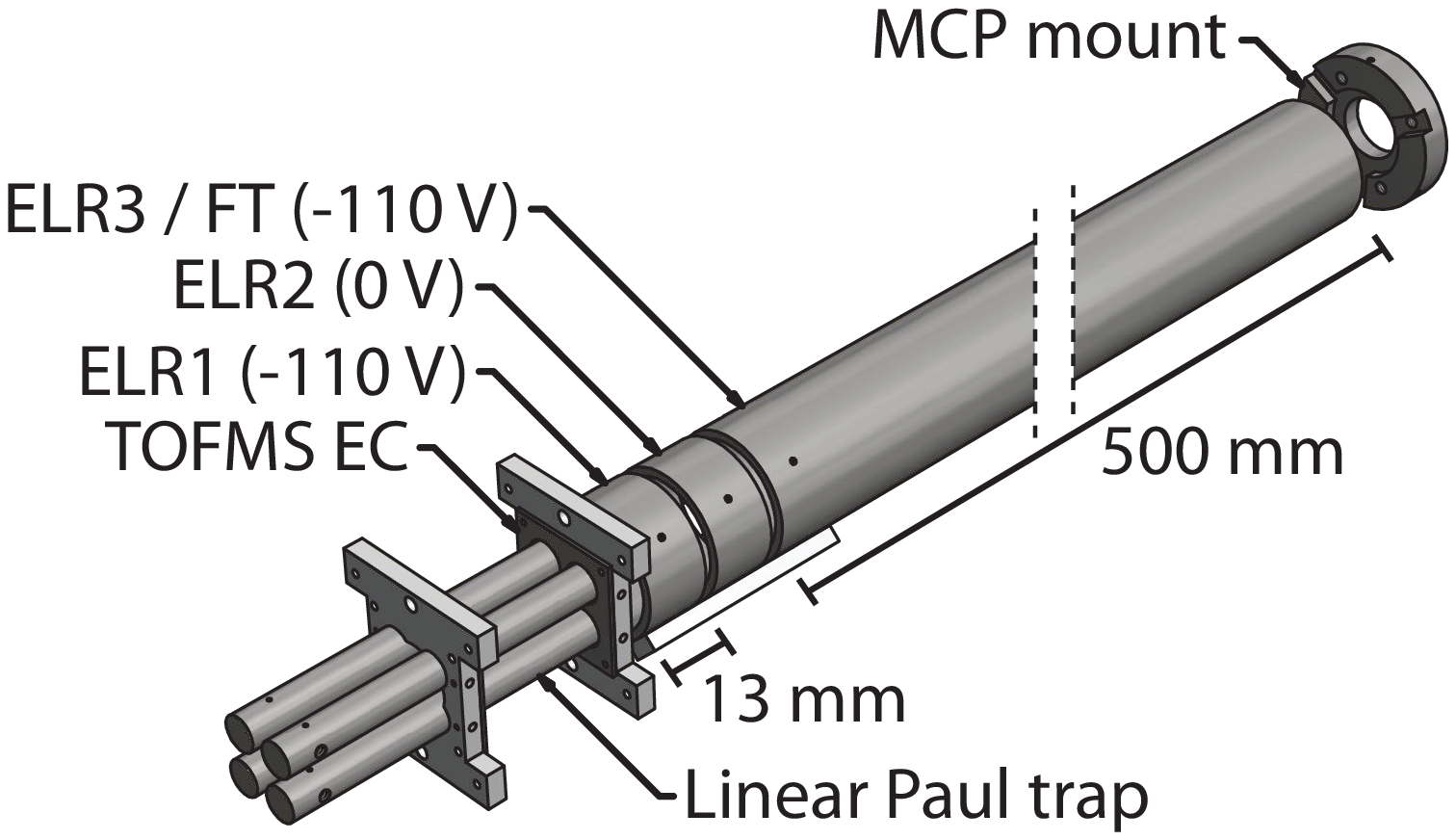}
  \caption{Schematic of the apparatus.  TOFMS EC labels the trap endcap that is switched at the start of TOFMS.  Adjacent to the TOFMS EC is the einzel lens, consisting of two rings (ELR1 and ELR2) with the third doubling as the flight tube (ELR3 / FT).}
  \label{apparatus}
\end{figure}

The experimental work is performed in a linear radiofrequency (RF) Paul trap (Fig.~\ref{apparatus}); r$_0$ = 3 mm, z$_0$ = 15 mm, $\Omega_{RF}/2\pi = 3.35$ MHz, $V_{pp} = 300$ V, with the RF applied to two opposing rod electrodes and the other two held at ground. The trap endcaps are plates with holes in the center, similar to a trap design shown in~\cite{tabor_suitability_2012,pedregosa_anharmonic_2010}, held at 850 V. Trapped ion species were identified and counted using TOFMS.  A 3-element Einzel lens utilizing a 500 mm flight tube as the last electrode spatially and temporally focuses ejected trapped ions onto a Hamamatsu F4655 multi-channel plate (MCP) operating at -3 kV with a -0.5 kV anode bias. The analog-mode MCP signal is terminated with $50\ \Omega$ at a digital oscilloscope with a sample rate of 2 GSa/s, triggered with the TOFMS initialization. To extract the ions, the TOFMS-side trap endcap is switched from 850 V to -110 V in 10 ns.  In order to guide ion extraction, the RF is left on for the beginning of the TOFMS sequence.  We find that the timing resolution is better if the RF is shut off early in the TOFMS sequence, possibly because of deleterious effects of fringe fields; we use a 10 $\mu$s delay before the RF is switched off with a 2 $\mu$s ring-down time.

\begin{figure}[htbp!]
  \centering
  \includegraphics*[width=3in]{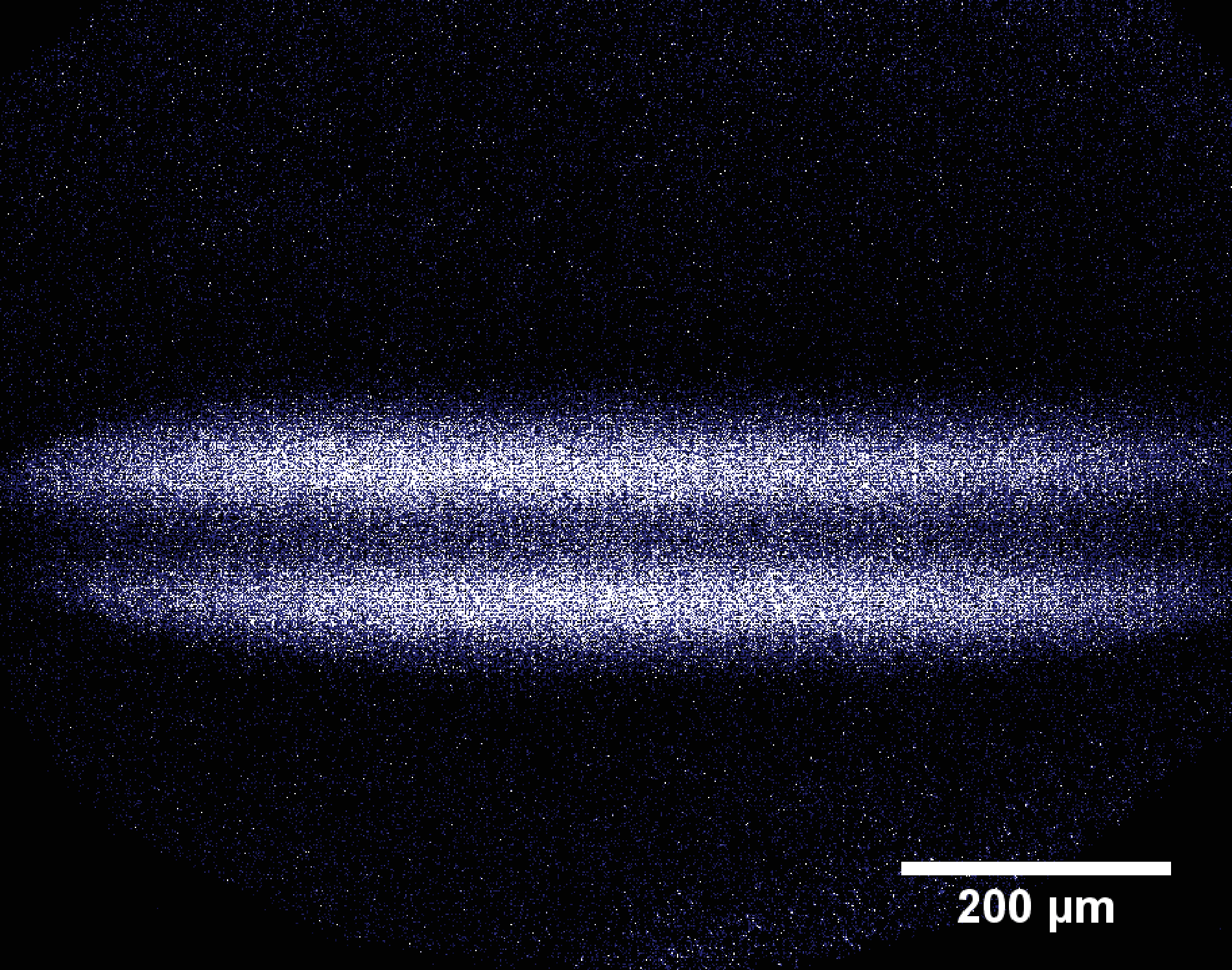}
  \caption{EMCCD bicrystal image.  The fluorescing ions are \Bae.  The dark core consists of other \Ba\ isotopes and \AlH.}
  \label{crystal}
\end{figure}

A typical experiment begins with ablation-loading and laser-cooling \Ba.  Al\+\ is then ablation-loaded and sympathetically cooled, forming a bicrystal as shown in \fig{crystal}.  The \Bae\ Doppler cooling beams enter the trap at approximately 30 degrees from the trap axis, resulting in the dark, non-fluorescing core being comprised of both other barium isotopes and Al\+.  Al\+\ then reacts with background gas to form \AlH; TOFMS and REMPD spectroscopy analysis determine that, under typical vacuum conditions with an N$_2$-calibrated ion gauge reading $1\times 10^{-10}$ Torr, the Al\+\ sample is fully converted to \AlH\ after a few minutes.  A non-evaporable getter pump and valve control the background gas pressure and hence reaction rate.

Rotational state analysis is accomplished destructively by ($1+1'$) REMPD with the two photon energies drawn in \fig{PEC}. In order for dissociation to proceed, the first photon must be resonant with an A-X bound-bound transition, and the target \AlH\ ion must be in the addressed rotational state. Since the kinetic energy released in dissociation is small compared with the trap depth, successful dissociation appears as conversion of \AlH\ to trapped Al\+. Thus, for fully efficient REMPD, the resulting ratio of Al\+\ to \AlH\ measures the fraction of \AlH\ initially in the probed rotational state.

\begin{figure}[htbp!]
  \centering
  \includegraphics*[width=3in]{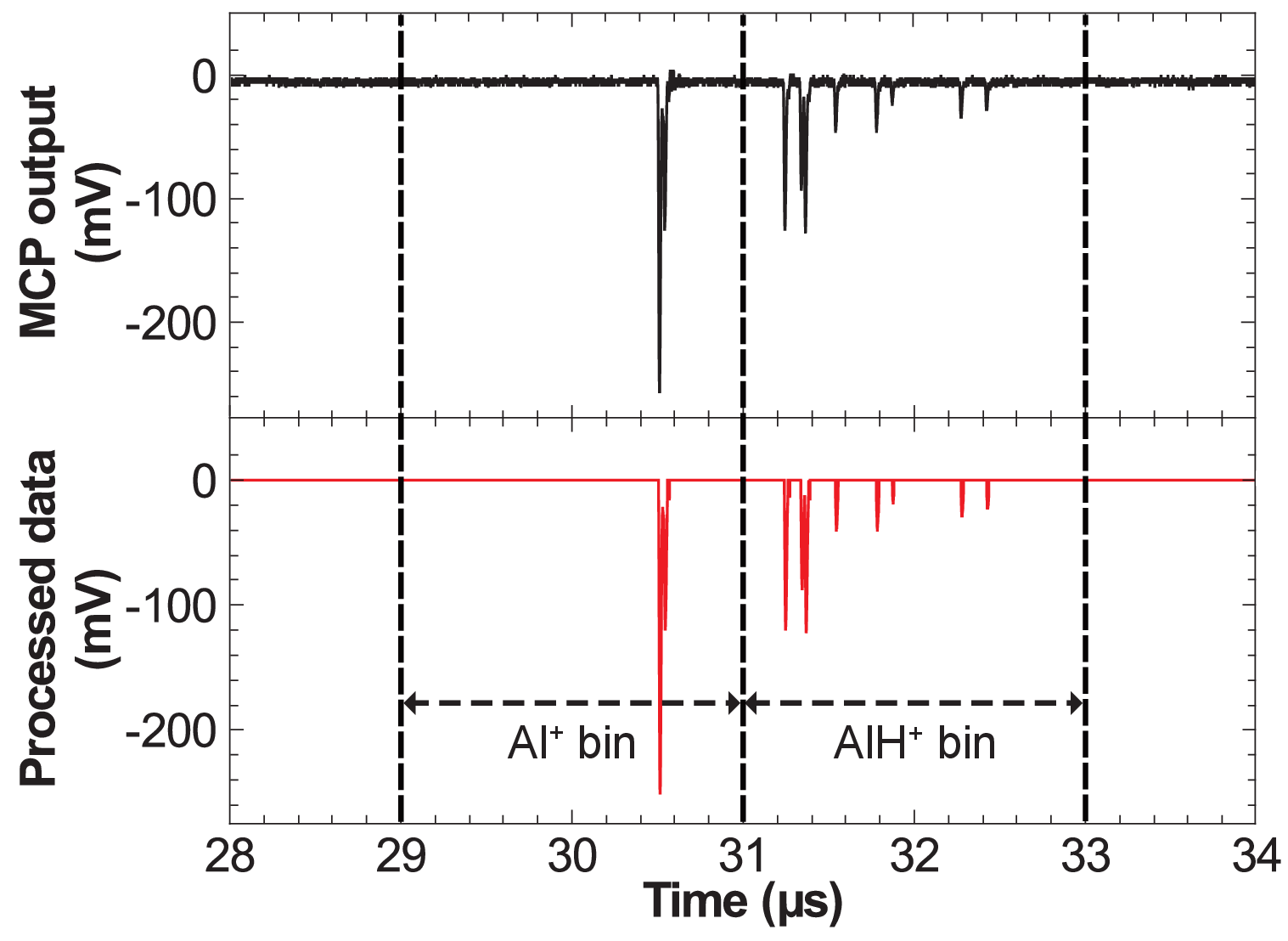}
  \caption{Sample TOFMS spectrum.  The upper plot (black) is the MCP output measured by an oscilloscope.  The lower plot (red) is the processed data with DC background and low-level noise removed.  For this spectrum, the TOFMS measured $13 \pm 4$ Al\+\ and $14 \pm 4$ AlH\+\ ions.  The single-species timing and the inter-species separation agree well with the values from a simulation. In spectra with a large Al\+\ fraction, as in the figure, we find that the timing dispersion of \AlH\ is increased compared to Al\+\ or pure \AlH\ samples.}
  \label{TOF}
\end{figure}

To quantify the number of ions detected by the MCP in each experimental run, the TOFMS oscilloscope trace is first processed by removing the DC background.  A discriminator level is then set to eliminate low-level background noise.  Each run, time windows of 2 $\mu$s on either side of a $t_0$ are assigned to Al\+\ and \AlH. Since the TOFMS timing varies over long timescales with patch-charging of electrodes, every experimental run is preceded by a TOFMS run with roughly 50\% of each species present, in order to optimally set $t_0$. Each signal window is numerically integrated to yield the total charge.  See \fig{TOF} for a typical data set before and after signal processing.  Histogramming the charge of isolated peaks in the processed data within the signal windows over multiple experimental runs yields a single ion charge of 8 pC, which is consistent with the MCP specifications.

In our experiment, we tune the 360 nm laser to different lines in the Q-branch of the \AlH\ A-X transition. Transitions originating from different X-state rotational levels are separated from each other by at least 160 GHz, so they are well resolved above the natural linewidth (typically 2 MHz) and the linewidth of the 360 nm laser (1.8 GHz). We reduce the laser intensity sufficiently to ensure that power broadening effects do not complicate the rotational state analysis.  The TOFMS sequence is triggered quickly following the REMPD pulses, much faster than the timescale of Al\+\ reaction with background gas.

\begin{figure}[htbp!]
  \centering
  \includegraphics*[width=3in]{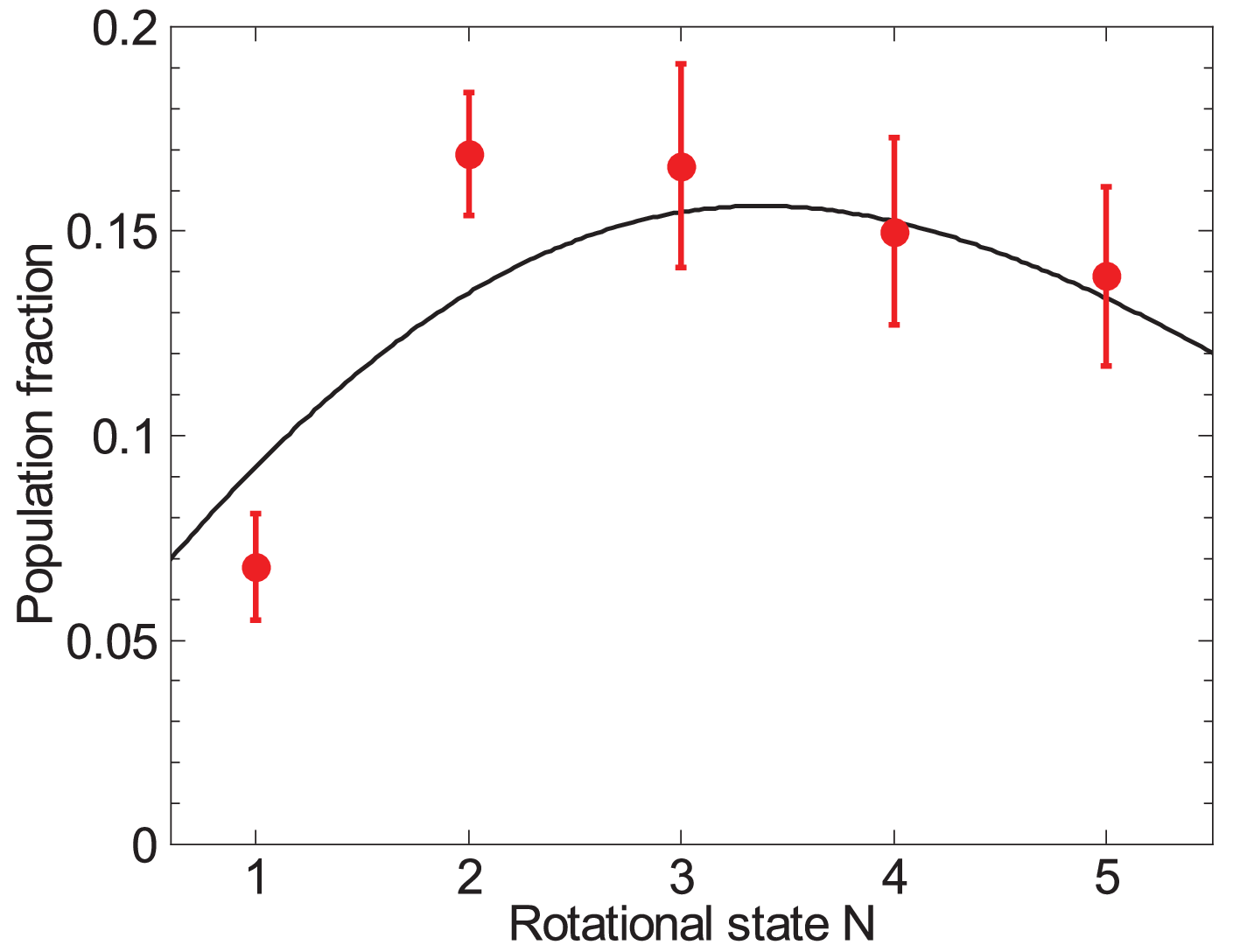}
  \caption{Thermal population distribution measured with REMPD/TOFMS.  The black line is the expected thermal distribution at 293 K, with no free parameters.  The red data points are the measured population in each rotational state with error bars from counting statistics.}
  \label{thermal}
\end{figure}

The REMPD 360 nm photon source is a frequency-doubled dye laser pumped by a Nd:YAG operating at the second harmonic; the average pulse energy is 50 $\mu$J with a spot size of 3.0 mm $\times$ 1.5 mm FWHM at trap center, providing sufficient intensity for \AlH\ ions to undergo 100 Rabi oscillations over a single pulse duration.  The 266 nm photon source is a second Nd:YAG operating at the fourth harmonic, synchronized and counterpropagating with the dye laser, with a pulse energy of 2.2 mJ and a spot size of 1000 x 270 $\mu$m FWHM at trap center.  The two REMPD pulses are fully overlapped temporally at trap center.  Alignment of the 266 nm light was performed by maximizing the rate of photoassisted $\textrm{Ba}^+ \rightarrow \textrm{Ba}^{2+}$. The mechanism for this slow but fortuitous process is not understood, but it possibly proceeds via the \Ba\ 8s~$^2\textrm{S}_{1/2}$ intermediate state which couples resonantly with a 264.8 nm photon to the \Ba\ 6p~$^2\textrm{P}_{1/2}$ state \cite{NIST_ASD}, which is excited during \Ba\ Doppler cooling.  Beam characterization is performed using a CMOS camera.

We use REMPD/TOFMS to map population of trapped \AlH\ in the first few rotational states (\fig{thermal}).  Since the \AlH\ production time is much longer than equilibration times with room temperature blackbody radiation, we expect a thermal distribution.  For those molecules in the probed rotational state, the calculated cross-section yields a 98\% single-shot \AlH\ dissociation probability for our pulse parameters; in our current implementation we apply 10 REMPD pulses per experiment. Several experimental runs probing a given rotational state are integrated to create the figure.  The agreement is good between our measured rotational distribution and the expected room temperature distribution.

\section{Conclusions}
Rotational state analysis is an essential ingredient for a wide range of applications using trapped molecular ions, including precision spectroscopy, quantum information processing, ultracold chemistry, and coherent control.  REMPD provides a destructive but powerful rotational state readout technique which can be relatively straightforward to implement in the laboratory.  Here, we have made necessary \emph{ab initio} structure calculations for \AlH\ and calculated the dissociation cross-section for a convenient REMPD pathway from the ground vibrational state.  The predicted cross-section is large enough to be readily accessible in the laboratory, and we perform the first experimental demonstration of ($1+1'$) REMPD of this species.  We use REMPD/TOFMS to analyze the rotational distribution of trapped \AlH\ and find consistency with the expected room temperature thermal distribution.

Using REMPD/TOFMS readout, we will be able to test a proposal to rotationally cool \AlH\ with a spectrally filtered pulsed laser~\cite{lien_optical_2011}.  The ability to quickly control rotational states with a single laser would make \AlH\ an especially interesting candidate for quantum information processing applications~\cite{demille_quantum_2002, schuster_cavity_2011} and precision searches for time-variation of the electron-proton mass ratio.  Since \AlH\ has a semi-cycling transition, it is also amenable to direct Doppler cooling  or to non-destructive state readout by fluorescence~\cite{nguyen_challenges_2011}.  For all future experiments with this species, REMPD/TOFMS analysis will either serve as the implemented state readout technique or as an essential starting point for developing non-destructive alternatives.

\section{Acknowledgments}
This work was supported by AFOSR, NSF, and the David and Lucile Packard Foundation.

\section*{References}
\bibliography{bibliography}

\end{document}